\documentclass[thmsa,11pt,letterpaper]{article}
\usepackage{amsfonts}
\usepackage{amsmath}

\input{tcilatex}

\begin{document}

\setcounter{page}{0} \topmargin0pt \oddsidemargin5mm \renewcommand{%
\thefootnote}{\fnsymbol{footnote}} \newpage \setcounter{page}{0} 
\begin{titlepage}
\begin{flushright}
EMPG-04-07 \\
\end{flushright}
\vspace{0.5cm}
\begin{center}
{\Large {\bf Auxiliary matrices on both sides of the equator} }

\vspace{0.8cm}
{ \large Christian Korff}

\vspace{0.5cm}
{\em School of Mathematics, University of Edinburgh\\
Mayfield Road, Edinburgh EH9 3JZ, UK}
\end{center}
\vspace{0.2cm}
 
\renewcommand{\thefootnote}{\arabic{footnote}}
\setcounter{footnote}{0}

\begin{abstract}
The spectra of previously constructed auxiliary matrices for the six-vertex model at roots of unity are 
investigated for spin-chains of even and odd length. The two cases show remarkable differences. In 
particular, it is shown that for even roots of unity and an odd number of sites the eigenvalues contain 
two linear independent solutions to Baxter's $TQ$-equation corresponding to the Bethe ansatz equations 
above and below the equator. In contrast, one finds for even spin-chains only one linear independent 
solution and complete strings. The other main 
result is the proof of a previous conjecture on the degeneracies of the six-vertex model at roots of unity. The 
proof rests on the derivation of a functional equation for the auxiliary matrices which is closely related to a 
functional equation for the eight-vertex model conjectured by Fabricius and McCoy.   
\end{abstract}
\vfill{ \hspace*{-9mm}
\begin{tabular}{l}
\rule{6 cm}{0.05 mm}\\
c.korff@ed.ac.uk
\end{tabular}}
\end{titlepage}\newpage

\section{Introduction}

Baxter's $TQ$-equation \cite{Bx71} is one of the corner-stones of integrable
systems and has been discussed in a variety of contexts. While it originated
from the Bethe ansatz computations for the six-vertex model, it provides
conceptually a more general framework to solve the transfer matrix
eigenvalue problem of integrable vertex models and paved the way for the
solution of the eight-vertex model \cite{Bx72,Bx73}. (See also \cite%
{FM8v,FM8v2} for recent developments.)

In the $TQ$-equation the $T$ symbolizes the transfer matrix of the
integrable model at hand and $Q$ is called the auxiliary matrix in terms of
which the transfer matrix can be expressed. To be concrete, consider the
six-vertex model on a square-lattice with periodic boundary conditions. Then
the $TQ$-equation is a second order linear difference equation of the form 
\begin{equation}
T(z)Q(z)=\phi (zq^{-2})Q(zq^{2})+\phi (z)Q(zq^{-2})  \label{TQ0}
\end{equation}
with $z\in \mathbb{C}$ being the spectral variable and $q$ the coupling (or
crossing) parameter of the model. The coefficients $\phi $ are given by some
known scalar function, which in our convention specified below will be $\phi
(z)=(zq^{2}-1)^{M}$. Here $M$ is the number of lattice columns. This
equation is often treated on different levels. Some authors interpret it
merely on the level of complex-valued functions, i.e. for the eigenvalues of
the respective matrices only, not addressing the usually harder problem of
the explicit construction of the matrix $Q$. Solving this construction
problem, however, allows one to address the problem of determining the
existence and number of possible solutions to the $TQ$-equation. This is
particularly important for those cases which one often finds to be
explicitly (or implicitly) excluded from the discussion, for instance when
the parameter $q$ is a root of unity or when the length of the spin-chain
associated with the vertex model is odd.\smallskip

For these \textquotedblleft special\textquotedblright\ cases the relation
between the spectrum of explicitly constructed $Q$-matrices in \cite%
{RW02,KQ,KQ2,KQ3} and the solutions of the $TQ$-equation will be our primary
interest in this article. For our discussion it will be important to
distinguish between the $TQ$-equation as a functional relation and an
operator equation.

\subsection{The $TQ$-equation in terms of functions.}

In \cite{KWLZ} Krichever et al pointed out that Baxter's $TQ$-equation
naturally appears as an auxiliary linear problem in the context of the
discretized Liouville equation, see Section 4 therein. For the six-vertex
model the analogue of this equation arises from the fusion hierarchy \cite%
{KR,Kun}, 
\begin{equation}
T^{(n)}(zq^{2})T^{(n)}(z)-T^{(n+1)}(z)T^{(n-1)}(zq^{2})=\phi (z)\phi
(zq^{2n})\ .  \label{Liouv}
\end{equation}
Here the $T^{(n)}$'s should be thought of as the eigenvalues of the
six-vertex fusion matrix of spin $n/2$ with $T^{(2)}=T$ being the transfer
matrix and $T^{(1)}=\phi $ the quantum determinant \cite{KRS}. We will give
the definition of the six-vertex fusion hierarchy below. The general
exposition given in \cite{KWLZ} (which also addresses the case of higher
rank and elliptic functions) concludes that there will be in general
two-linear independent solutions, say $\mathcal{Q}^{\pm }$, to the
functional equation (\ref{TQ0}) which satisfy the Wronskian 
\begin{equation}
\mathcal{Q}^{+}(zq^{2})\mathcal{Q}^{-}(z)-\mathcal{Q}^{+}(z)\mathcal{Q}%
^{-}(zq^{2})=\phi (z)\ .  \label{wronski0}
\end{equation}
From this Wronskian two types of ``complementary''\ Bethe ansatz equations
can be deduced \cite{KWLZ}. Notice that after equation (4.27) in \cite{KWLZ}
the restriction to elliptic polynomials $\phi $ of even degree is made which
corresponds to the case of spin-chains of even length. While the focus in 
\cite{KWLZ} is mainly on the elliptic case, one wonders about the
trigonometric limit and the implications for the six-vertex model.

Pronko and Stroganov investigated this question of two potential linear
independent solutions of the six-vertex $TQ$-equation in \cite{PrSt}.
Starting with the XXX spin-chain their discussion is generalized to the XXZ
spin-chain excluding the root of unity case. They identify for spin-chains
with an odd number of sites $M$ one solution, say $\mathcal{Q}^{+}$, as the
familiar solution from the Bethe ansatz, while the second solution $\mathcal{%
Q}^{-}$ is associated with solving the Bethe ansatz on the \textquotedblleft
wrong side of the equator\textquotedblright\ meaning that it incorporates $%
M-n$ Bethe roots. Here $0\leq n\leq M/2$ is the number of down-spins in the
corresponding eigenstate of the transfer matrix or spin-chain Hamiltonian.
For spin-chains with an even number of sites, however, there appears to be
only one solution in the case of periodic boundary conditions. Baxter
addresses this phenomenon on the basis of the coordinate Bethe ansatz and
numerical computations in \cite{Bx02} stating that the missing Bethe roots
for the second solution $\mathcal{Q}^{-}$ have gone off to infinity. (See
also the third paragraph after equation (4.43) in \cite{KWLZ}.) When a
nonzero horizontal electric field is applied, numerical computations show
both linear independent solutions exist for $M$ even and odd \cite{Bx02}.

The difference between spin-chains of odd and even length has been further
underlined in \cite{Odd}. For the special root of unity case $q^{3}=1$ and
spin-chains with an odd number of sites the two independent solutions of the
six-vertex $TQ$-equation have been explicitly constructed \cite{Odd}
starting from a conjecture on the particularly simple form of the
groundstate \cite{Bx72a,Bx89,ABB88}.\smallskip

Our discussion in this article will show that the solutions obtained at $%
q^{3}=1$ in \cite{Odd} are in fact the \emph{only} linear independent
solutions in the spin $S^{z}=\pm 1/2$ sector which possess the expected
number of $(M-1)/2$ and $(M+1)/2$ Bethe roots, respectively. This is due to
the large degeneracies in the spectrum of the transfer matrix connected with
the loop algebra symmetry of the six-vertex model at roots of unity \cite%
{DFM}. A similar reduction in the number of solutions takes place at higher
roots of unity as well. In other words, the number and nature of solutions
to the $TQ$-equation (\ref{TQ0}) does also crucially depend on the value of
the parameter $q$ and not only on the length of the spin-chain.\smallskip

The various cases outlined and the necessary distinctions one has to make
clearly show the importance of the explicit construction of auxiliary
matrices in order to obtain better control over the different scenarios.
However, Baxter's construction of auxiliary matrices for the six and
eight-vertex model as presented in his book \cite{BxBook} applies only to
spin-chains of even length (see the comment after equation (9.8.16)) and
also at roots of unity can only be extended to a particular subset of cases.
In contrast, the auxiliary matrices constructed in \cite{KQ} for roots of
unity $q^{N}=1$ and the ones in \cite{RW02} for the case of
\textquotedblleft generic\textquotedblright\ $q$ do not have such
limitations and apply for even as well as odd length of the chain. They are
also of a simpler algebraic form. In this work we will explicitly relate
those auxiliary matrices' spectrum to the two linear independent solutions
of the $TQ$-equation by extending the discussion of \cite{KQ2} from the even
to the odd case.

\subsection{The $TQ$-equation in terms of operators.}

Starting with the papers \cite{BLZ97,BLZ99} on the Liouville model and
subsequent papers on the six-vertex model \cite{AF97,RW02,KQ} a new approach
to construct auxiliary matrices has been developed which relies on
representation theory. In this method one first solves the Yang-Baxter
equation to obtain a matrix which commutes with the transfer matrix and
afterwards derives the $TQ$-equation by investigating the decomposition of
certain tensor products of representations. The novel feature \cite{RW02,KQ}
in this context is the appearance of additional free parameters,
collectively called $p,$ in the auxiliary matrix, i.e. $Q=Q(z;p)$. These
free parameters shift in the operator solution of the $TQ$-equation \cite%
{RW02,KQ}, 
\begin{equation}
T(z)Q(z;p)=\phi (zq^{-2})Q(zq^{2};p^{\prime })+\phi (z)Q(zq^{-2};p^{\prime
\prime })\ .
\end{equation}
Thus, in this construction method one has to consider a generalized version
of Baxter's $TQ$-equation at the level of operators. The free parameters
are, for instance, necessary to break spin-reversal symmetry or to lift the
degeneracies of the transfer matrix at roots of unity \cite{KQ}. They also
contain information on the analytic structure of the eigenvalues of the
auxiliary matrices \cite{KQ2,KQ3}.

For the case when $q$ is not a root of unity and twisted boundary conditions
the spectra of of the auxiliary matrices constructed in \cite{RW02} have
been computed using the algebraic Bethe ansatz \cite{KQ3}. It was found that
the eigenvalues decompose into two parts which are related by spin-reversal 
\cite{KQ3}, 
\begin{equation}
Q(z;p)=Q^{+}(z;p)Q^{-}(z;p)\ .  \label{Q0}
\end{equation}
For special choices of the parameters $p$ the functions $Q^{\pm }$ are the
eigenvalues of the lattice analogues of the $Q$-operators constructed by
Bazhanov et al for the Liouville model \cite{BLZ97,BLZ99}. For twisted
boundary conditions these two parts of the eigenvalue (\ref{Q0}) can indeed
be identified (up to some normalization constants) with the two linear
independent solutions $\mathcal{Q}^{\pm }$ of the $TQ$-equation discussed in
the previous section, see \cite{KQ3} and the conclusions of this article.
However, in general this is not true, since the two linear independent
solutions $\mathcal{Q}^{\pm }$ might not always exist, as for example in the
case of periodic boundary conditions and spin-chains of even length. Notice
also that Baxter's $Q$-operator for even spin-chains corresponds to only one
of these eigenvalue parts, say $Q^{+}$. This should make clear that the $Q$%
-operators in \cite{RW02,KQ} not only differ in the construction procedure
but are quite different objects from Baxter's $Q$ discussed in \cite{BxBook}.

\subsection{Outline and results of this article}

The outline of the article is as follows:

\begin{description}
\item[Section 2.] We introduce the fusion hierarchy of the six-vertex model
and fix our conventions. The fusion matrices are defined such that they are
polynomials of maximal degree $M$ in the spectral variable $z$. Here $M$ is
the length of the associated XXZ spin-chain.

\item[Section 3.] We recall the definition of a particular subset of the
auxiliary matrices at roots of unity constructed in \cite{KQ} and discuss
the general form of their spectra defining the subparts $Q^{\pm }$ of the
eigenvalue decomposition (\ref{Q0}). Preliminary results have already been
obtained in \cite{KQ2} for the case of spin-chains with $M$ even. Additional
results for $M$ even and odd are contained in \cite{KQ3}. In this article we
will complete our investigation of the spectra by extending the discussion
of \cite{KQ2} to spin-chains with an odd number of sites.

\item[Section 4.] The eigenvalues of the auxiliary matrices are in general
polynomials in the spectral variable. In this section we determine when they
have maximal degree and when they vanish at the origin. Both facts are
related to the absence or occurrence of infinite Bethe roots.

\item[Section 5.] We review from \cite{KQ2} the discussion of the $TQ$%
-equation in terms of the eigenvalues of the auxiliary matrices. In
particular, we discuss the transformation under spin-reversal and show how
the second linear independent solution to the $TQ$-equation arises.

\item[Section 6.] One of the main results of this article is the derivation
of a functional equation, for $M$ even and odd, that proves a previously
formulated conjecture in \cite{KQ2}, see equations (18) and (19) therein.
This result will enable us to determine the level of degeneracy of the
eigenvalues of the transfer matrix at roots of unity and relate the two
parts $Q^{\pm }$ of the eigenvalues via an inversion formula. Moreover, this
result can be considered as a six-vertex analogue of the eight-vertex
functional equation conjectured by Fabricius and McCoy in \cite{FM8v,FM8v2}.
We will comment on this in the conclusions.

\item[Section 7.] The results for the case of odd spin-chains will show that
for even roots of unity and periodic boundary conditions we obtain both
linear independent solutions $\mathcal{Q}^{\pm }$ of the $TQ$-equation from
the auxiliary matrices in \cite{KQ}. This is a constructive existence proof
for these solutions. For odd roots of unity we will see that the eigenstates
of the transfer matrix associated with $\mathcal{Q}^{\pm }$ (in the case
that both solutions exist) correspond to zero eigenvalues of the auxiliary
matrices. We in particular make contact with Stroganov's solutions for $N=3$ 
\cite{Odd} and show that they are the only ones with the expected number of
Bethe roots (see also the appendix).

\item[Section 8.] We state our conclusions and relate our results to the
case when $q$ is not a root of unity considered in \cite{KQ3} and the recent
developments in the eight-vertex model \cite{FM8v,FM8v2}.
\end{description}

\section{The six-vertex fusion hierarchy}

We introduce the six-vertex model from a representation theoretic point of
view. Denote by $\pi _{z}^{(n)}:U_{q}(\widetilde{sl}_{2})\rightarrow 
\limfunc{End}\mathbb{C}^{n+1}$ the spin $n/2$ evaluation representation of
the quantum loop algebra, i.e. 
\begin{eqnarray}
\pi _{z}^{(n)}(e_{1})\left\vert m\right\rangle &=&[n-m+1]_{q}\left\vert
m-1\right\rangle ,\quad \pi _{z}^{(n)}(f_{0})=z^{-1}\pi _{z}^{(n)}(e_{1}), 
\notag \\
\pi _{z}^{(n)}(f_{1})\left\vert m\right\rangle &=&[m+1]_{q}\left\vert
m+1\right\rangle ,\quad \pi _{z}^{(n)}(e_{0})=z\pi _{z}^{(n)}(f_{1}),  \notag
\\
\pi _{z}^{(n)}(q^{h_{1}})\left\vert m\right\rangle &=&q^{n-2m}\left\vert
m\right\rangle ,\quad \pi _{z}^{(n)}(q^{h_{0}})=\pi _{z}^{(n)}(q^{-h_{1}}),
\label{pin}
\end{eqnarray}
with $m=0,1,...,n$. Define the fusion matrix of degree $n+1$ by setting 
\begin{equation}
T^{(n+1)}(zq^{-n-1})=\limfunc{Tr}_{0}L_{0M}^{(n+1)}(zq^{n+1})\cdots
L_{01}^{(n+1)}(zq^{-n-1})  \label{Tn}
\end{equation}
where $L^{(n+1)}$ is the intertwiner with respect to the tensor product $\pi
_{w}^{(n)}\otimes \pi _{1}^{(1)}$, 
\begin{eqnarray}
\left\langle 0\right\vert L^{(n+1)}(w)\left\vert 0\right\rangle &=&wq\,\pi
^{(n)}(q^{h_{1}/2})-\pi ^{(n)}(q^{-h_{1}/2}),\quad  \notag \\
\left\langle 0\right\vert L^{(n+1)}(w)\left\vert 1\right\rangle
&=&wq\,(q-q^{-1})\pi ^{(n)}(q^{h_{1}/2})\pi ^{(n)}(f_{1}),  \notag \\
\left\langle 1\right\vert L^{(n+1)}(w)\left\vert 0\right\rangle
&=&(q-q^{-1})\pi ^{(n)}(e_{1})\pi ^{(n)}(q^{-h_{1}/2}),\quad  \notag \\
\left\langle 1\right\vert L^{(n+1)}(w)\left\vert 1\right\rangle &=&wq\,\pi
^{(n)}(q^{-h_{1}/2})-\pi ^{(n)}(q^{h_{1}/2})\ .  \label{fusL2}
\end{eqnarray}
Here the scalar products are taken in the second factor of the tensor
product, i.e. the spin 1/2 representation. The fusion matrices satisfy the
functional equation \cite{KR} 
\begin{equation}
T^{(n)}(z)T^{(2)}(zq^{-2})=\left( zq^{2}-1\right)
^{M}T^{(n+1)}(zq^{-2})+\left( z-1\right) ^{M}T^{(n-1)}(zq^{2})\;.
\label{fus}
\end{equation}
The fusion hierarchy contains two special elements from which all others can
be successively generated, namely the six-vertex transfer matrix\footnote{%
Note that our definition of the six-vertex transfer matrix differs from the
one in \cite{KQ,KQ2,KQ3} by an overall factor, 
\begin{equation*}
T(z)\rightarrow q^{\frac{M}{2}}T(z)/(zq^{2}-1)^{M}\ .
\end{equation*}%
} $T$ and the quantum determinant $T^{(1)}$ \cite{KRS} which are obtained
via the identification 
\begin{equation}
T^{(2)}(zq^{-2})\equiv T(z)\quad \quad \text{and}\quad \quad
T^{(1)}(z)\equiv (zq^{2}-1)^{M}\;.
\end{equation}
In this manner the above functional equation may also serve as a defining
relation for the fusion matrices. An alternative form of the fusion
hierarchy is the one given in the introduction, see (\ref{Liouv}). Both
versions are equivalent. From the transfer matrix $T$ we obtain as
logarithmic derivative the $XXZ$ spin-chain Hamiltonian, 
\begin{eqnarray}
H_{\text{XXZ}} &=&-\left( q-q^{-1}\right) ~\left. z\frac{d}{dz}\ln \frac{T(z)%
}{(zq^{2}-1)^{M}}\right\vert _{z=1} \\
&=&-\frac{1}{2}\sum_{m=1}^{M}\left\{ \sigma _{m}^{x}\sigma _{m+1}^{x}+\sigma
_{m}^{y}\sigma _{m+1}^{y}+\frac{q+q^{-1}}{2}\,(\sigma _{m}^{z}\sigma
_{m+1}^{z}-1)\right\} \;.
\end{eqnarray}
The well-known symmetries of the model are expressed in terms of the
following commutators 
\begin{equation}
\lbrack T^{(m)}(z),T^{(n)}(w)]=[T^{(n)}(z),S^{z}]=[T^{(n)}(z),\mathfrak{R}%
]=[T^{(n)}(z),\mathfrak{S}]=0\;,  \label{symm}
\end{equation}
where the respective operators are defined as 
\begin{equation}
S^{z}=\frac{1}{2}\sum_{m=1}^{M}\sigma _{m}^{z},\text{\quad }\mathfrak{R}%
=\sigma ^{x}\otimes \cdots \otimes \sigma ^{x}=\prod\limits_{m=1}^{M}\sigma
_{m}^{x},\text{\quad }\mathfrak{S}=\sigma ^{z}\otimes \cdots \otimes \sigma
^{z}=\prod\limits_{m=1}^{M}\sigma _{m}^{z}\;.  \label{SzRS}
\end{equation}
These symmetries hold for spin-chains of even as well as odd length.

\section{Auxiliary matrices at roots of unity}

In a series of papers \cite{KQ,KQ2,KQ3} auxiliary matrices for the
six-vertex model at roots of unity have been constructed. The construction
procedure and the difference with Baxter's method have been discussed in 
\cite{KQ} and we refer the reader to this work for details. In order to keep
this paper self-contained we briefly give the definition of a special
subvariety of the set of auxiliary matrices constructed in \cite{KQ}.

\subsection{Definition}

Suppose $q$ is a primitive root of unity of order $N$ and set $N^{\prime }=N$
if the order is odd and $N^{\prime }=N/2$ if it is even. Define the
following $N^{\prime }$-dimensional nilpotent evaluation representation $\pi
_{w}^{\mu }$ of the quantum loop algebra $U_{q}(\widetilde{sl}_{2})$ \cite%
{RA89,CK}, 
\begin{eqnarray}
\pi ^{\mu }(q^{h_{1}})\left\vert n\right\rangle &=&\mu
^{-1}q^{-2n-1}\left\vert n\right\rangle ,\quad \pi ^{\mu }(f_{1})\left\vert
n\right\rangle =\left\vert n+1\right\rangle \;,\quad \pi ^{\mu
}(f_{1})\left\vert N^{\prime }-1\right\rangle =0\;,  \notag \\
\pi ^{\mu }(e_{1})\left\vert n\right\rangle &=&\frac{\mu +\mu ^{-1}-\mu
q^{2n}-\mu ^{-1}q^{-2n}}{(q-q^{-1})^{2}}\;\left\vert n-1\right\rangle ,
\label{pimu}
\end{eqnarray}
and 
\begin{equation}
\pi _{w}^{\mu }(q^{h_{0}})=\pi ^{\mu }(q^{-h_{1}}),\quad \pi _{w}^{\mu
}(f_{0})=w^{-1}\pi ^{\mu }(e_{1}),\quad \pi _{w}^{\mu }(e_{0})=w~\pi ^{\mu
}(f_{1}),\quad w,\mu \in \mathbb{C}^{\times }\;.
\end{equation}
Let the matrix 
\begin{equation}
L^{\mu }=\left( 
\begin{array}{cc}
\alpha _{\mu } & \beta _{\mu } \\ 
\gamma _{\mu } & \delta _{\mu }%
\end{array}
\right) =\alpha _{\mu }\otimes \sigma ^{+}\sigma ^{-}+\beta _{\mu }\otimes
\sigma ^{+}+\gamma _{\mu }\otimes \sigma ^{-}+\delta _{\mu }\otimes \sigma
^{-}\sigma ^{+}\;.  \label{Ldec}
\end{equation}
be the intertwiner of the tensor product $\pi _{w}^{\mu }\otimes \pi
_{z=1}^{(1)}$ of evaluation representations, explicitly 
\begin{eqnarray}
\alpha _{\mu }(w) &=&wq\,\pi ^{\mu }(q^{h_{1}/2})-\pi ^{\mu
}(q^{-h_{1}/2}),\quad \beta _{\mu }(w)=wq(q-q^{-1})\pi ^{\mu
}(q^{h_{1}/2})\pi ^{\mu }(f_{1}),\;  \notag \\
\gamma _{\mu } &=&\left( q-q^{-1}\right) \pi ^{\mu }(e_{1})\pi ^{\mu
}(q^{-h_{1}/2}),\quad \delta _{\mu }(w)=wq\,\pi ^{\mu }(q^{-h_{1}/2})-\pi
^{\mu }(q^{h_{1}/2})\ .  \label{Lmu}
\end{eqnarray}
Define the auxiliary matrix in terms of these matrices as the trace of the
following operator product, 
\begin{equation}
Q_{\mu }(z)=\limfunc{Tr}_{0}L_{0M}^{\mu }(z/\mu )L_{0M-1}^{\mu }(z/\mu
)\cdots L_{01}^{\mu }(z/\mu )\ .  \label{Qmu}
\end{equation}
This matrix commutes by construction with the fusion matrices, 
\begin{equation}
\lbrack Q_{\mu }(w),T^{(n)}(z)]=0,  \label{QT0}
\end{equation}
and preserves two of the symmetries (\ref{symm}) \cite{KQ,KQ2}, 
\begin{equation}
\lbrack Q_{\mu }(z),S^{z}]=[Q_{\mu }(z),\mathfrak{S}]=0\ .  \label{QSS}
\end{equation}
Spin-reversal symmetry on the other hand is broken \cite{KQ,KQ2}, 
\begin{equation}
\mathfrak{R}Q_{\mu }(z,q)\mathfrak{R}=Q_{\mu ^{-1}}(z\mu ^{-2},q)=Q_{\mu
^{-1}}(zq^{2}\mu ^{-2},q^{-1})^{t}=(-zq/\mu )^{M}Q_{\mu }(z^{-1}q^{-2}\mu
^{2})^{t}\ .  \label{RQR}
\end{equation}
These relations hold for all $M$ and allow one to determine the conjugate
transpose of the auxiliary matrix \cite{KQ2}, 
\begin{equation}
Q_{\mu }(z,q)^{\ast }=Q_{\bar{\mu}}(\bar{z},q^{-1})^{t}=Q_{\bar{\mu}}(\bar{z}%
q^{-2},q)\ .  \label{Qad}
\end{equation}
In addition, one derives from the following non-split exact sequence of
evaluation representations $\pi _{w}^{\mu }\;$\cite{KQ} 
\begin{equation}
0\rightarrow \pi _{w^{\prime }}^{\mu q}\hookrightarrow \pi _{w}^{\mu
}\otimes \pi _{z}^{(1)}\rightarrow \pi _{w^{\prime \prime }}^{\mu
q^{-1}}\rightarrow 0,\quad w=w^{\prime }q^{-1}=w^{\prime \prime }q=z/\mu
\label{seq1}
\end{equation}
the $TQ$-equation 
\begin{equation}
T(z)Q_{\mu }(z)=\left( z-1\right) ^{M}~Q_{\mu
q}(zq^{2})+(zq^{2}-1)^{M}~Q_{\mu q^{-1}}(zq^{-2})\ .
\end{equation}
The proof can be found in \cite{KQ}, here we will only review parts of the
calculation of the spectrum of the auxiliary matrices given in \cite{KQ2}
and extend the results therein to spin-chains of odd length.

\subsection{The general form of the spectrum}

The starting point is the same as in \cite{KQ2}: provided that the
commutation relation 
\begin{equation}
\lbrack Q_{\mu }(z),Q_{\nu }(w)]=0,\quad \quad \mu ,\nu ,z,w\in \mathbb{C}
\label{assumption}
\end{equation}
holds, the eigenvectors of $Q_{\mu }(z)$ are independent of the parameter $%
\mu $ as well as the spectral variable $z$. In order to prove (\ref%
{assumption}) one has to explicitly construct the corresponding intertwiners
of the tensor products $\pi _{w}^{\mu }\otimes \pi _{1}^{\nu }$ for all $%
N^{\prime }\in \mathbb{N}$. As pointed out in \cite{KQ2} the necessary
conditions for these intertwiners to exist are satisfied for all roots of
unity. An explicit construction has been carried out for $N=3$ \cite{KQ2}
and $N=6$. Numerical checks have been performed for $N^{\prime }=4,5,7$. We
shall take this as sufficient evidence for (\ref{assumption}) to hold true.

There are two important implications of (\ref{assumption}). The first is
that the auxiliary matrices are normal and hence diagonalizable, see (\ref%
{Qad}). The second consequence is that the eigenvalues of $Q_{\mu }$ must be
polynomial in the spectral variable $z$. Their most general form is
therefore given by\footnote{%
Throughout this paper we will denote eigenvalues and operators by the same
symbol.} 
\begin{eqnarray}
Q_{\mu }(z) &=&\mathcal{N}_{\mu }\;z^{n_{\infty }}P_{B}(z)P_{\mu
}(z)P_{S}(z^{N^{\prime }},\mu ^{2N^{\prime }})  \notag \\
&=&\mathcal{N}_{\mu }\;z^{n_{\infty
}}\prod_{i=1}^{n_{+}}(1-z/z_{i})\prod_{i=1}^{n_{-}}(1-z/w_{i}(\mu
))\prod_{i=1}^{n_{S}}(1-z^{N^{\prime }}/a_{i}(\mu ))\;.  \label{QE}
\end{eqnarray}
Note that we slightly differ in the notation from \cite{KQ2} and have
redefined the normalization constant $\mathcal{N}_{\mu }$ by setting $%
P_{B}(0)=P_{\mu }(0)=P_{S}(0)=1$. Besides these minor differences our
definition of the various polynomials entering the eigenvalues is the same
as in \cite{KQ2}.

\begin{itemize}
\item The monomial factor in front of the eigenvalue is related to the
occurrence of vanishing and infinite Bethe roots when the root of unity
limit is taken in the deformation parameter $q$.

\item The second factor $P_{B}$ contains only roots $z_{i}$ which do not
depend on the free parameter $\mu $ and which will be identified with the
finite Bethe roots at roots of unity. Moreover, we exclude from this set
complete or exact strings, i.e. for every $z_{i}$ there is at least one
integer $0<\ell <N^{\prime }$ such that $z_{i}q^{2\ell }$ is not a root of $%
P_{B}$.

\item The third factor contains roots which do depend on the parameter $\mu $%
. For even chains, $M\in 2\mathbb{N}$, this factor was identified in \cite%
{KQ2} with the rescaled polynomial $P_{B},$%
\begin{equation*}
P_{\mu }(z)=P_{B}(z\mu ^{-2})\ .
\end{equation*}
Here we will find that this relation ceases to be valid for even roots of
unity when $M\in 2\mathbb{N}+1$. Again we exclude the possibility that the
roots of $P_{\mu }$ occur in strings.

\item Finally, the last factor $P_{S}$ contains all roots which occur in
strings. The zeroes $a_{i}$ may or may not depend on the parameter $\mu $.
Because all roots are sitting in a string the polynomial depends on $%
z^{N^{\prime }}$ rather than $z$. Note that we allow for the possibility $%
n_{S}=0$.
\end{itemize}

\noindent For later purposes let us decompose the eigenvalues of the
auxiliary matrices into two parts similar as it has been done in \cite%
{BLZ99,KQ3} for $q$ \textquotedblleft generic\textquotedblright . Namely, we
set 
\begin{equation}
Q^{+}(z)=P_{B}(z)=\prod_{i=1}^{n_{+}}(1-z/z_{i})  \label{PB}
\end{equation}
and secondly, 
\begin{equation}
Q^{-}(z)=\lim_{\mu \rightarrow q^{-N^{\prime }}}\mathcal{N}_{\mu
}\;z^{n_{\infty }}P_{\mu }(z)P_{S}(z^{N^{\prime }},\mu ^{2N^{\prime }})\;.
\label{Qm}
\end{equation}
This decomposition may seem arbitrary at the moment but it will become clear
in our line of argument. Notice that we have eliminated the dependence of
the auxiliary matrix on the free parameter $\mu $ in (\ref{Qm}). Moreover,
it is worth stressing that $Q^{\pm }$ are not always both identical with the
two linear independent solutions $\mathcal{Q}^{\pm }$ of the $TQ$-equation
(the latter might even not exist), 
\begin{equation}
T(z)\equiv T^{(2)}(zq^{-2})=(zq^{2}-1)^{M}q^{\mp S^{z}}\frac{\mathcal{Q}%
^{\pm }(zq^{-2})}{\mathcal{Q}^{\pm }(z)}+(z-1)^{M}q^{\pm S^{z}}\frac{%
\mathcal{Q}^{\pm }(zq^{2})}{\mathcal{Q}^{\pm }(z)}
\end{equation}
which satisfy the Wronskian 
\begin{equation}
q^{S^{z}}\mathcal{Q}^{+}(zq^{2})\mathcal{Q}^{-}(z)-q^{-S^{z}}\mathcal{Q}%
^{+}(z)\mathcal{Q}^{-}(zq^{2})=(q^{S^{z}}-q^{-S^{z}})(1-zq^{2})^{M}\ .
\label{wronski}
\end{equation}
The additional phase factors $q^{\pm S^{z}}$ in comparison with the equation
(\ref{wronski0}) of \cite{KWLZ} discussed in the introduction are due to
different conventions. We shall set $\mathcal{Q}^{+}(0)=\mathcal{Q}^{-}(0)=1$
and $\deg \mathcal{Q}^{+}=M-\deg \mathcal{Q}^{-}=M/2-S^{z}$. While $Q^{+}$
is by definition the solution $\mathcal{Q}^{+}$ above the equator, $%
Q^{-}\neq \mathcal{Q}^{-}$ in general. In particular we have included the
normalization constant $\lim_{\mu \rightarrow q^{-N^{\prime }}}\mathcal{N}%
_{\mu }$ in the definition of $Q^{-}$, which in some cases can be zero as we
will discuss below.

\section{The degree of the eigenvalues and \textquotedblleft
infinite\textquotedblright\ Bethe roots}

We start our discussion with the first factor in the eigenvalue (\ref{QE}),
the monomial $z^{n_{\infty }}$ which is related to the fact that some Bethe
roots in the root of unity limit vanish or tend to infinity. Obviously, $%
n_{\infty }\neq 0$ if and only if the eigenvalue of the auxiliary matrices
vanishes at the origin. Another obvious observation is that by construction
of the auxiliary matrices it follows that 
\begin{equation}
\deg Q_{\mu }=n_{\infty }+n_{+}+n_{-}+n_{S}N^{\prime }\leq M\;.  \label{deg}
\end{equation}
As it turns out this upper bound is assumed if and only if we have a
vanishing monomial contribution, i.e. $n_{\infty }=0$. This can be deduced
from the following relation for the auxiliary matrices \cite{KQ,KQ2} 
\begin{equation}
Q_{\mu }(z)=(-zq/\mu )^{M}Q_{\mu ^{-1}}(z^{-1}q^{-2})^{t}
\end{equation}
which implies that the coefficients in the power series expansion 
\begin{equation}
Q_{\mu }(z)=\sum_{m=0}^{M}Q_{\mu }^{(m)}z^{m}
\end{equation}
are related via 
\begin{equation}
Q_{\mu }^{(m)}=(-\mu )^{-M}q^{-M+2m}\left( Q_{\mu ^{-1}}^{(M-m)}\right)
^{t}\;.
\end{equation}
In particular, setting $m=0$ in the above identity we see that the
eigenvalue of $Q_{\mu }$ is of degree $M$ whenever $Q_{\mu }(0)=Q_{\mu
}^{(0)}\neq 0$. At the same time this clearly prevents $n_{\infty }\neq 0$.
The coefficient $Q_{\mu }^{(0)}$ can be easily calculated by noting that the
building blocks $L^{\mu }(z)$ of the auxiliary matrix are lower triangular
matrices at $z=0$, 
\begin{equation}
L^{\mu }(0)=-\pi ^{\mu }(q^{-h_{1}/2})\otimes \sigma ^{+}\sigma ^{-}-\pi
^{\mu }(q^{h_{1}/2})\otimes \sigma ^{-}\sigma ^{+}+(q-q^{-1})\pi ^{\mu
}(e_{1}q^{-h_{1}/2})\otimes \sigma ^{-}\;.
\end{equation}
Thus, the matrix $Q_{\mu }(0)$ in quantum space is diagonal with its
diagonal elements given by 
\begin{equation}
Q_{\mu }(0)=(-)^{M}\limfunc{Tr}_{\pi ^{\mu }}q^{-h_{1}S^{z}}=(-)^{M}(\mu
q)^{S^{z}}\sum_{\ell =0}^{N^{\prime }-1}q^{2\ell S^{z}}\;.
\end{equation}
For the interpretation of this result we distinguish the following cases:

\begin{enumerate}
\item When $q^{N^{\prime }}=1$, i.e. for primitive roots of unity of odd
order, the degree of the polynomial will only be equal to $M$ in the
commensurate sectors $2S^{z}=0\func{mod}N$. At the same time this means that
there are no infinite Bethe roots, that is $n_{\infty }=0$. Consequently, we
have 
\begin{equation}
Q_{\mu }(0)=\mathcal{N}_{\mu }=(-)^{M}N~q^{S^{z}}\mu ^{S^{z}},\qquad
q^{N^{\prime }}=1,\;2S^{z}=0\func{mod}N\ .  \label{N1}
\end{equation}

\item When $q^{N^{\prime }}=-1$ we have to distinguish between $M$ even and
odd. Let $M$ be even then $S^{z}$ takes integer values only and we obtain $%
\deg Q_{\mu }=M$ if and only if $2S^{z}=0\func{mod}N.$ Again we find in
these spin-sectors the normalization constant 
\begin{equation}
Q_{\mu }(0)=\mathcal{N}_{\mu }=N^{\prime }q^{S^{z}}\mu ^{S^{z}},\qquad
q^{N^{\prime }}=-1,\;M\in 2\mathbb{N},\;2S^{z}=0\func{mod}N\ .  \label{N2}
\end{equation}

\item For $q^{N^{\prime }}=-1$ and $M$ odd, however, the total spin
eigenvalue will be constrained to the set $2S^{z}\in 2\mathbb{Z}+1$,
preventing the existence of a monomial factor. That is, in this case we
always have $\deg Q_{\mu }=M$ and $n_{\infty }=0$. The normalization
constant is therefore 
\begin{equation}
Q_{\mu }(0)=\mathcal{N}_{\mu }=-q^{N^{\prime }S^{z}}\mu ^{S^{z}}\ \frac{%
q^{N^{\prime }S^{z}}-q^{-N^{\prime }S^{z}}}{q^{S^{z}}-q^{-S^{z}}},\quad
\quad q^{N^{\prime }}=-1,\;M\in 2\mathbb{N}+1\ .  \label{N3}
\end{equation}
\end{enumerate}

\section{The $TQ$-equation}

The most important property of the auxiliary matrix is the solution of the
following functional equation with the six-vertex transfer matrix which has
been proved to hold for $M$ even and odd \cite{KQ}, 
\begin{equation}
T(z)Q_{\mu }(z)=\left( z-1\right) ^{M}~Q_{\mu
q}(zq^{2})+(zq^{2}-1)^{M}~Q_{\mu q^{-1}}(zq^{-2})\ .
\end{equation}
From this functional equation and the fact that $T^{(2)}$ does not depend on
the free parameter $\mu $ we infer similar to the case $M$ even considered
in \cite{KQ2} that 
\begin{equation*}
w_{i}(\mu )=w_{i}\mu ^{2}\qquad \text{and}\qquad a_{i}(\mu )=a_{i}(\mu
^{2N^{\prime }})
\end{equation*}
implying the following form for the eigenvalues of the transfer matrix, 
\begin{equation}
T(z)=\frac{\mathcal{N}_{\mu q}}{\mathcal{N}_{\mu }}~q^{2n_{\infty }}\left(
z-1\right) ^{M}~\frac{Q^{+}(zq^{2})}{Q^{+}(z)}+\frac{\mathcal{N}_{\mu q^{-1}}%
}{\mathcal{N}_{\mu }}~q^{-2n_{\infty }}(zq^{2}-1)^{M}~\frac{Q^{+}(zq^{-2})}{%
Q^{+}(z)}\ .  \label{TQp}
\end{equation}
Here the ratios of the normalization constants can only depend on $q$ from
which we deduce 
\begin{equation}
\frac{\mathcal{N}_{\mu q}}{\mathcal{N}_{\mu }}=\frac{\mathcal{N}_{\mu }}{%
\mathcal{N}_{\mu q^{-1}}}\ .
\end{equation}
The zeroes of the polynomial $Q^{+}$ are fixed through the \textquotedblleft
Bethe ansatz\textquotedblright\ equations, 
\begin{equation}
0=(1-z_{i}q^{2})^{M}q^{-s}Q^{+}(z_{i}q^{-2})+(1-z_{i})^{M}q^{s}Q^{+}(z_{i}q^{2}),\qquad i=1,...,n_{+},
\label{BAE}
\end{equation}
with 
\begin{equation}
q^{s}:=\frac{\mathcal{N}_{\mu q}}{\mathcal{N}_{\mu }}q^{2n_{\infty }}=\frac{%
\mathcal{N}_{\mu }}{\mathcal{N}_{\mu q^{-1}}}q^{2n_{\infty }}\ .  \label{s}
\end{equation}
We will argue below that this phase factor is determined by the total spin
of the eigenstate and the number of Bethe roots which tend to zero in the
root of unity limit.

\subsection{Spin reversal and Bethe roots \textquotedblleft beyond the
equator\textquotedblright}

We are now exploring the role of the polynomial factor $P_{\mu }$. Following
the same line of argument as in \cite{KQ2} we act with the spin reversal
operator from both sides on the $TQ$-equation employing the transformation
law \cite{KQ,KQ2} 
\begin{equation*}
\mathfrak{R}Q_{\mu }(z)\mathfrak{R}=Q_{\mu ^{-1}}(z\mu ^{-2})
\end{equation*}
of the auxiliary matrix. Replacing afterwards $\mu \rightarrow \mu ^{-1}$ we
obtain the equation 
\begin{equation}
T(z)Q_{\mu }(z\mu ^{2})=\left( z-1\right) ^{M}~Q_{\mu q^{-1}}(z\mu
^{2})+(zq^{2}-1)^{M}~Q_{\mu q}(z\mu ^{2})\ .  \label{TQm}
\end{equation}
Since we employed the spin-reversal operator we refer to this identity as
the $TQ$-equation ``beyond the equator''. The corresponding expression in
terms of eigenvalues is deduced to be 
\begin{eqnarray}
T(z) &=&\frac{\mathcal{N}_{\mu q^{-1}}}{\mathcal{N}_{\mu }}~\left(
z-1\right) ^{M}\frac{P_{\mu q^{-1}}(z\mu ^{2})}{P_{\mu }(z\mu ^{2})}+\frac{%
\mathcal{N}_{\mu q}}{\mathcal{N}_{\mu }}~\left( zq^{2}-1\right) ^{M}\frac{%
P_{\mu q}(z\mu ^{2})}{P_{\mu }(z\mu ^{2})}  \notag \\
&=&\left( z-1\right) ^{M}q^{-s}~\frac{Q^{-}(zq^{2})}{Q^{-}(z)}%
+(zq^{2}-1)^{M}q^{s}~\frac{Q^{-}(zq^{-2})}{Q^{-}(z)}\ .  \label{TQ2}
\end{eqnarray}
Notice that in comparison with (\ref{TQp}) not only the phase factors $q^{s}$
have been inverted but that the polynomial $Q^{-}$ will in general have a
different degree than $Q^{+}$, i.e. $n_{-}\neq n_{+}$. From (\ref{TQm}) we
now obtain the ``Bethe ansatz equations beyond the equator'', 
\begin{equation}
0=(1-w_{i}q^{2})^{M}q^{s}Q^{-}(w_{i}q^{-2})+(1-w_{i})^{M}q^{-s}Q^{-}(w_{i}q^{2}),\quad i=1,...,n_{-}\ .
\label{BAE2}
\end{equation}
This second solution to the $TQ$-equation, which is related by spin-reversal
to $Q^{+}$, will not always be linear independent as we shall see below. The
parameter $s$ entering the phase factors in the eigenvalue expressions of
the transfer matrices changes sign, since we already saw for particular
cases, see (\ref{N1}), (\ref{N2}), (\ref{N3}), that it is related it to the
total spin of the corresponding eigenvectors through the normalization
constant. For the general case its value can be determined by making contact
with the fusion hierarchy.

Using the functional equation (\ref{fus}) for the fusion matrices presented
in the introduction, the above results for the transfer matrix are extended
to all fusion matrices via induction. A straightforward calculation yields 
\begin{equation}
T^{(n)}(z)=q^{\pm (n+1)s}Q^{\pm }(z)Q^{\pm }(zq^{2n})\sum_{\ell =1}^{n}\frac{%
q^{\mp 2\ell s}(zq^{2\ell }-1)^{M}}{Q^{\pm }(zq^{2\ell })Q^{\pm }(zq^{2\ell
-2})}\ .  \label{fusQpm}
\end{equation}
In the paper \cite{KQ3} it has been argued, using the algebraic Bethe ansatz
when $q$ is not a root of unity, that the parameter $s$ can be identified
with 
\begin{equation}
s=2n_{0}+S^{z}\func{mod}N^{\prime }\ .
\end{equation}
Here $n_{0}$ denotes the number of Bethe roots which vanish in the root of
unity limit $q^{N}\rightarrow 1$.

\section{A functional equation relating $Q^{+}$ and $Q^{-}$}

The final step in the analysis of the spectrum of the auxiliary matrices
rests on the following functional equation, which has been proved for $N=3$
in \cite{KQ2}, 
\begin{equation}
Q_{\mu }(z\mu ^{2}q^{2})Q_{\nu }(z)=(zq^{2}-1)^{M}Q_{\mu \nu q}(z\mu
^{2}q^{2})+q^{N^{\prime }M}Q_{\mu \nu q^{-N^{\prime }+1}}(z\mu
^{2}q^{2})T^{(N^{\prime }-1)}(zq^{2})\;.  \label{QQQ}
\end{equation}
This equation is a direct consequence of the following decomposition of the
tensor product $\pi _{w}^{\mu }\otimes \pi _{1}^{\nu }$ of evaluation
representations, 
\begin{equation}
0\rightarrow \pi _{w^{\prime }}^{\mu ^{\prime }}\hookrightarrow \pi
_{w}^{\mu }\otimes \pi _{1}^{\nu }\rightarrow \pi _{w^{\prime \prime }}^{\mu
^{\prime \prime }}\otimes \pi _{z^{\prime }}^{(N^{\prime }-2)}\rightarrow 0
\label{seq2}
\end{equation}
where 
\begin{equation}
w=\mu \nu q^{2},\quad \mu ^{\prime }=\mu \nu q,\quad w^{\prime }=\mu q,\quad
\mu ^{\prime \prime }=\mu \nu q^{-N^{\prime }+1},\quad w^{\prime \prime
}=\mu q^{-N^{\prime }+1},\quad z^{\prime }=\nu q^{N^{\prime }+1}\;.
\end{equation}
For the moment assume the functional equation (\ref{QQQ}) to hold, the
derivation of (\ref{QQQ}) and (\ref{seq2}) is given in the appendix. Let us
insert the explicit form of the eigenvalue into (\ref{QQQ}). We find 
\begin{multline}
\frac{Q_{\mu }(z\mu ^{2}q^{2})Q_{\nu }(z)}{Q_{\mu \nu q}(z\mu ^{2}q^{2})}= 
\notag \\
\frac{\mathcal{N}_{\mu }\mathcal{N}_{\nu }}{\mathcal{N}_{\mu \nu q}}%
\,z^{n_{\infty }}P_{B}(z)P_{\mu =1}(zq^{2})\prod_{i=1}^{n_{S}}\frac{%
(1-z^{N^{\prime }}\mu ^{2N^{\prime }}/a_{i}(\mu ^{2N^{\prime
}}))(1-z^{N^{\prime }}/a_{i}(\nu ^{2N^{\prime }}))}{(1-z^{N^{\prime }}\mu
^{2N^{\prime }}/a_{i}(\mu ^{2N^{\prime }}\nu ^{2N^{\prime }})}=  \notag \\
\frac{\mathcal{N}_{\mu \nu q^{N^{\prime }+1}}}{\mathcal{N}_{\mu \nu q}}%
\;q^{N^{\prime }M}T^{(N^{\prime }-1)}(zq^{2})+(zq^{2}-1)^{M}  \label{eq}
\end{multline}
Here we have exploited the fact that the zeroes of the factors $P_{\mu
},\,P_{\nu }$\ and $P_{\mu \nu q^{N^{\prime }+1}}$, $P_{\mu \nu q}$ only
depend on $\mu ^{2},\nu ^{2}$ and $\mu ^{2}\nu ^{2}q^{2}$, respectively. The
possible complete string contribution $P_{S}$ contains the various
parameters only to the power $2N^{\prime }$. Notice that the last line of
the above equation is independent of the free parameters $\mu ,\nu $ (the
ratio of the normalization constants only depends on $q$ as pointed out
earlier). This implies that the zeroes $a_{i}(\mu ^{2N^{\prime }})$ are
either independent of $\mu $ altogether or only incorporate it as a
multiplicative factor, i.e. one has (exactly as in the case $N=3$ proved in 
\cite{KQ2}) the alternative 
\begin{equation}
a_{i}(\mu ^{2N^{\prime }})=a_{i}\quad \quad \text{or}\quad \text{\quad }%
a_{i}(\mu ^{2N^{\prime }})=a_{i}~\mu ^{2N^{\prime }}\;.  \label{ai}
\end{equation}
Hence, there are at most $2^{n_{S}}$ possible eigenvalues of the auxiliary
matrix in a degenerate eigenspace of the transfer matrix with fixed $%
n_{\infty }$ and $n_{\pm }$. This proves part of the second conjecture made
in \cite{KQ2}, see equation (18) with $a_{i}\equiv (z_{i}^{S})^{N^{\prime }}$%
. From the outcome on the zeroes of the complete string contribution one
deduces that the factor originating from the $P_{S}$ polynomials simplifies, 
\begin{equation}
\prod_{i=1}^{n_{S}}\frac{(1-z^{N^{\prime }}\mu ^{2N^{\prime }}/a_{i}(\mu
^{2N^{\prime }}))(1-z^{N^{\prime }}/a_{i}(\nu ^{2N^{\prime }}))}{%
(1-z^{N^{\prime }}\mu ^{2N^{\prime }}/a_{i}(\mu ^{2N^{\prime }}\nu
^{2N^{\prime }})}=P_{S}(z^{N^{\prime }},\mu =1)\;.
\end{equation}
Hence, we can rewrite the functional equation in terms of $Q^{\pm }$ as
follows, 
\begin{eqnarray*}
\frac{Q_{\mu }(z\mu ^{2}q^{2})Q_{\nu }(z)}{Q_{\mu \nu q}(z\mu ^{2}q^{2})} &=&%
\frac{\mathcal{N}_{\mu }}{\mathcal{N}_{\mu q^{-N^{\prime }+1}}}%
\,q^{-2n_{\infty }}Q^{+}(z)Q^{-}(zq^{2}) \\
&=&\frac{\mathcal{N}_{\mu q^{2N^{\prime }+1}}}{\mathcal{N}_{\mu
q^{-N^{\prime }+1}}}\;q^{N^{\prime }M}T^{(N^{\prime
}-1)}(zq^{2})+(zq^{2}-1)^{M}\;.
\end{eqnarray*}
Here we have set $\nu =q^{-N^{\prime }}$ in the normalization constants
without loss of generality. Invoking now the earlier stated form of the
fusion matrices (\ref{fusQpm}) this identity becomes 
\begin{multline*}
\frac{\mathcal{N}_{\mu }}{\mathcal{N}_{\mu q^{-N^{\prime }+1}}}%
\,q^{-2n_{\infty }}Q^{+}(z)Q^{-}(zq^{2})=\frac{\mathcal{N}_{\mu q}}{\mathcal{%
N}_{\mu q^{-N^{\prime }+1}}}q^{-s}Q^{+}(z)Q^{-}(zq^{2})= \\
\frac{\mathcal{N}_{\mu q^{2N^{\prime }+1}}}{\mathcal{N}_{\mu q^{-N^{\prime
}+1}}}\;q^{N^{\prime }M\pm N^{\prime }s}Q^{\pm }(z)Q^{\pm
}(zq^{2})\sum_{\ell =1}^{N^{\prime }-1}\frac{q^{\mp 2\ell s}(zq^{2\ell
+2}-1)^{M}}{Q^{\pm }(zq^{2\ell +2})Q^{\pm }(zq^{2\ell })}+(zq^{2}-1)^{M}\;.
\end{multline*}
Setting $\mu \rightarrow q^{-N^{\prime }}$ and solving the last expression
for $Q^{\pm }$ we obtain the result 
\begin{equation}
Q^{\mp }(z)=\,q^{\pm (N^{\prime }+1)s}Q^{\pm }(z)\sum_{\ell =1}^{N^{\prime }}%
\frac{q^{-2\ell s}(zq^{2\ell }-1)^{M}}{Q^{\pm }(zq^{2\ell })Q^{\pm
}(zq^{2\ell -2})}\ .  \label{inv}
\end{equation}
Hence, the two solutions of the $TQ$-equation are related to each other by
an inversion formula (provided that $Q^{-}\neq 0$, see the discussion
below). This identity is the six-vertex analogue of the functional equation
conjectured by Fabricius and McCoy for Baxter's 1972 auxiliary matrix of the
eight-vertex model \cite{FM8v,FM8v2}. (We will comment further on this in
the conclusion.) Here we have proved this functional equation in the
six-vertex limit for all roots of unity and spin-chains of even as well as
odd length. Let us investigate the difference between the solutions $Q^{\pm
} $ depending on the cases when $M$ is even or odd.

\paragraph{Decomposition of the eigenvalue for $M$ even.}

For spin-chains of even length $M=2m$ and at $q^{N^{\prime }}=\pm 1$ the sum
in (\ref{Qm}), which at first sight appears to be a rational function,
simplifies due to the Bethe ansatz equations (\ref{BAE}) to a polynomial,
i.e. we have for any contour $C$ encircling the point $z_{i}q^{-2\ell
^{\prime }},$%
\begin{multline*}
\frac{1}{2\pi i}\oint\limits_{C(z_{i}q^{-2\ell ^{\prime }})}\sum_{\ell
=1}^{N^{\prime }}\frac{q^{-2\ell s}(zq^{2\ell }-1)^{M}}{Q^{+}(zq^{2\ell
})Q^{+}(zq^{2\ell -2})}\ dz= \\
\frac{q^{-2\ell ^{\prime }s}(z_{i}-1)^{M}}{Q^{+}(z_{i}q^{-2})\tprod%
\limits_{j\neq i}(1-z_{i}/z_{j})}+\frac{q^{-2(\ell ^{\prime
}+2)s}(z_{i}q^{2}-1)^{M}}{Q^{+}(z_{i}q^{2})\tprod\limits_{j\neq
i}(1-z_{i}/z_{j})}=0,\qquad 0<\ell ^{\prime }\leq N^{\prime }\ .
\end{multline*}
As a consequence we obtain in general the identifications \cite{KQ2} 
\begin{equation}
\lim_{\mu \rightarrow q^{-N^{\prime }}}P_{\mu }(z)=Q^{+}(z)  \label{identity}
\end{equation}
and 
\begin{equation}
\lim_{\mu \rightarrow q^{-N^{\prime }}}\mathcal{N}_{\mu }\;z^{n_{\infty
}}P_{S}(z^{N^{\prime }},\mu ^{2N^{\prime }})=q^{(N^{\prime }+1)s}\sum_{\ell
=1}^{N^{\prime }}\frac{q^{-2\ell s}(zq^{2\ell }-1)^{M}}{Q^{+}(zq^{2\ell
})Q^{+}(zq^{2\ell -2})}\;.  \label{PS}
\end{equation}
We therefore conclude that $Q^{-}\neq \mathcal{Q}^{-}$ in this case. In
fact, one finds numerically that the solution $\mathcal{Q}^{-}$ with $%
M/2+S^{z}$ Bethe roots does not exist. The other important conclusion is
that the result (\ref{PS}) together with (\ref{ai}) enables us to read off
the degeneracy of the transfer matrix eigenvalue corresponding to $Q^{+}$.
According to (\ref{ai}) each zero $a_{i}$ in the complete string
contribution $P_{S}$ is either independent of $\mu $ or is multiplied by a
factor $\mu ^{2N^{\prime }}$ showing that there are $2^{n_{S}}$ possible
eigenvalues of the auxiliary matrices each corresponding to a vector in the
degenerate eigenspace of the transfer matrix. This is in agreement with the
observation \cite{DFM} that only spin-1/2 representations occur in the
tensor products describing the finite-dimensional representations of the
loop algebra. Obviously, if $n_{S}=0$ the eigenvalue of the transfer matrix
is non-degenerate up to spin-reversal symmetry.

\paragraph{Decomposition of the eigenvalue for $M$ odd.}

For $M=2m+1$ odd the above simplification of the eigenvalue in general also
holds true when $N$ odd with the possible exception that the part (\ref{PS})
of the eigenvalue completely vanishes, we will discuss this case below. For $%
N$ even, however, the polynomials $P_{\mu }$ and $P_{B}=Q^{+}$ differ. Since
we have now $s\in \frac{1}{2}\mathbb{Z}$ and $q^{N^{\prime }}=-1$ the
rational function (\ref{PS}) has poles at $z=z_{i}$ with non-vanishing
residue, 
\begin{multline*}
\frac{1}{2\pi i}\oint\limits_{C(z_{i})}\sum_{\ell =1}^{N^{\prime }}\frac{%
q^{-2\ell s}(1-zq^{2\ell })^{M}}{Q^{+}(zq^{2\ell })Q^{+}(zq^{2\ell -2})}~dz=
\\
\frac{q^{-2s}(z_{i}q^{2}-1)^{M}}{Q^{+}(z_{i}q^{2})\tprod\limits_{j\neq
i}(1-z_{i}/z_{j})}+\frac{q^{-2N^{\prime }s}(z_{i}-1)^{M}}{%
Q^{+}(z_{i}q^{-2})\tprod\limits_{j\neq i}(1-z_{i}/z_{j})}\neq 0
\end{multline*}
Therefore, the factor $Q^{+}(z)$ in front of the sum is needed to cancel
these poles and the above factorization (\ref{identity}), (\ref{PS}) does
not take place. As a consequence complete strings are absent and there are
no additional degeneracies other than spin-reversal symmetry. As argued
earlier in Section 4 infinite Bethe roots are absent as well and we must
have, 
\begin{equation}
Q^{-}(z)=\lim_{\mu \rightarrow q^{-N^{\prime }}}\mathcal{N}_{\mu }~P_{\mu
}(z)=q^{(N^{\prime }+1)S^{z}}Q^{+}(z)\sum_{\ell =1}^{N^{\prime }}\frac{%
q^{-2\ell S^{z}}(zq^{2\ell }-1)^{M}}{Q^{+}(zq^{2\ell })Q^{+}(zq^{2\ell -2})}
\end{equation}
with the degrees of the polynomials $Q^{\pm }$ obeying 
\begin{equation}
\deg Q^{-}=M-\deg Q^{+}=\frac{M}{2}+S^{z}\ .
\end{equation}
Thus, we obtain a very different picture depending on the length of the
spin-chain being odd or even.

\paragraph{The quantum Wronskian.}

The difference between the two situations of even and odd spin-chains is
highlighted further by introducing the analogue of \ (\ref{wronski}) for the
two different parts of the auxiliary matrix eigenvalues. This corresponds to
the \textquotedblleft quantum Wronskian\textquotedblright\ in \cite{BLZ99}.
First note that using (\ref{inv}) we easily obtain 
\begin{equation}
q^{ns}Q^{+}(zq^{2n})Q^{-}(z)-q^{-ns}Q^{+}(z)Q^{-}(zq^{2n})=(q^{N^{\prime
}s}-q^{-N^{\prime }s})T^{(n)}(z)\ .
\end{equation}
Upon specializing to $n=1$ this relation simplifies to 
\begin{equation}
q^{s}Q^{+}(zq^{2})Q^{-}(z)-q^{-s}Q^{+}(z)Q^{-}(zq^{2})=(q^{N^{\prime
}s}-q^{-N^{\prime }s})(zq^{2}-1)^{M}\ .  \label{wronski3}
\end{equation}
Notice that the right hand of the above equation always vanishes except for
odd spin-chains and even roots of unity. This signals the linear dependence
between $Q^{\pm }$ for $M$ even and $q^{N^{\prime }}=1,\ M$ odd as described
above, compare with (\ref{identity}) and (\ref{PS}).

For $M$ odd and $q^{N^{\prime }}=-1$ the quantum Wronskian is non-zero and
we can identify 
\begin{equation}
Q^{+}=\mathcal{Q}^{+}\qquad \text{and\qquad }Q^{-}=\mathcal{N}_{\mu
=q^{-N^{\prime }}}\mathcal{Q}^{-}=-\frac{q^{N^{\prime }S^{z}}-q^{-N^{\prime
}S^{z}}}{q^{S^{z}}-q^{-S^{z}}}\ \mathcal{Q}^{-}\ .  \label{id}
\end{equation}
Thus, via an explicit construction of diagonalizable $Q$-operators we have
shown existence of the solutions above and below the equator. Notice that
the Wronskian implies the Bethe ansatz equations (\ref{BAE}). Namely, we
have for each zero $z_{i}$ of $Q^{+}$ that 
\begin{equation}
q^{S^{z}}\frac{Q^{+}(z_{i}q^{2})}{(z_{i}q^{2}-1)^{M}}=\frac{q^{N^{\prime
}S^{z}}-q^{-N^{\prime }S^{z}}}{Q^{-}(z_{i})}=-q^{S^{z}}\frac{%
Q^{+}(z_{i}q^{-2})}{(z_{i}-1)^{M}}\ .
\end{equation}
An analogous relation holds for the zeroes $w_{i}$ of $Q^{-}$ leading to the
Bethe ansatz equations beyond the equator (\ref{BAE2}). Note that (\ref%
{wronski}), respectively (\ref{wronski3}), contains more information than
each copy of the Bethe ansatz equations by itself, as it relates the zeroes $%
z_{i}$ and $w_{i}$ through the following sum rules for each $0\leq m\leq M$, 
\begin{equation}
\binom{M}{m}=\sum_{k+\ell =m}\frac{q^{S^{z}-2\ell }e_{k}^{+}e_{\ell
}^{-}-q^{-S^{z}-2k}e_{k}^{+}e_{\ell }^{-}}{q^{S^{z}}-q^{-S^{z}}}\ .
\label{wronski2}
\end{equation}
Here we have in light of (\ref{id}) identified the zeroes of $\mathcal{Q}%
^{\pm }$ with $\{z_{i}\}$ and $\{w_{i}\}$ and introduced the elementary
symmetric polynomials 
\begin{equation}
e_{k}^{+}=e_{k}(z_{1}^{-1},...,z_{n_{+}}^{-1})\qquad \text{and}\qquad
e_{k}^{-}=e_{k}(w_{1}^{-1},...,w_{M-n_{+}}^{-1})\ .  \label{epm}
\end{equation}
Numerically it is by far more feasible to solve this set of $M$ equations,
which is \emph{quadratic} in the $M$ variables $\{e_{k}^{+}\}\cup
\{e_{k}^{-}\}$, rather than the original $n_{+}=M/2-S^{z}$ Bethe ansatz
equations (\ref{BAE}) which are of order $M$ in the $n_{+}$ variables $%
\{z_{i}\}$. We verified for $N^{\prime }=3,5$ up to spin-chains of length $%
M=11$ that the number of solutions of the equations (\ref{wronski2}) matches
the dimension of the respective eigenspaces of the transfer matrix.

\section{Zero eigenvalues at $N$ and $M$ odd}

An additional aspect in which the cases of even and odd spin-chains differ
is the occurrence of zero eigenvalues of the auxiliary matrices. That the
auxiliary matrices can have indeed a non-trivial kernel for odd $M$ has
already been remarked upon in \cite{KQ} where it was noted for the simple
case of the $M=3$ spin-chain. As we will see it is closely connected with
the inversion formula (\ref{inv}) which follows from the functional equation
(\ref{QQQ}) and the two independent solutions $\mathcal{Q}^{\pm }$ of the $%
TQ $-equation.

\paragraph{Eigenstates with a maximum number of Bethe roots.}

Suppose $q$ is an odd root of unity then, as discussed above, the rational
function (\ref{PS}) becomes a polynomial in $z$ and $P_{\mu }(z)=Q^{+}(z\mu
^{-2})$. The only exception to this scenario is the case when the
corresponding eigenstate of the transfer matrix is a singlet. According to
our previous discussion, we therefore must have $2^{n_{S}}=1$, i.e. the
corresponding eigenvalue of the auxiliary matrix cannot contain complete
strings. In the absence of infinite Bethe roots we now argue that this
implies (\ref{PS}) vanishes. This can be deduced in several ways. Suppose
the two linear independent solutions $\mathcal{Q}^{\pm }$ to the $TQ$%
-equation exist. Then they have to obey the quantum Wronskian (\ref{wronski}%
). Solving the latter for $\mathcal{Q}^{-}$ we obtain 
\begin{equation}
\mathcal{Q}^{-}(z)=q^{-2S^{z}}\frac{\mathcal{Q}^{+}(z)\mathcal{Q}^{-}(zq^{2})%
}{\mathcal{Q}^{+}(zq^{2})}+(1-q^{-2S^{z}})\frac{(1-zq^{2})^{M}}{\mathcal{Q}%
^{+}(zq^{2})}\ .
\end{equation}
Iteration of this formula yields after $N$-steps 
\begin{equation*}
\mathcal{Q}^{-}(z)=q^{-2NS^{z}}\frac{\mathcal{Q}^{+}(z)\mathcal{Q}%
^{-}(zq^{2N})}{\mathcal{Q}^{+}(zq^{2N})}+(q^{2S^{z}}-1)\mathcal{Q}%
^{+}(z)\sum_{\ell =1}^{N}\frac{q^{-2\ell S^{z}}(1-zq^{2\ell })^{M}}{\mathcal{%
Q}^{+}(zq^{2\ell })\mathcal{Q}^{+}(zq^{2\ell -2})}
\end{equation*}
which upon invoking the root of unity condition $q^{N^{\prime }}=q^{N}=1$
gives 
\begin{equation}
\sum_{\ell =1}^{N}\frac{q^{-2\ell S^{z}}(zq^{2\ell }-1)^{M}}{\mathcal{Q}%
^{+}(zq^{2\ell })\mathcal{Q}^{+}(zq^{2\ell -2})}=0\ .  \label{zeroPS}
\end{equation}
This fact together with the identification $Q^{+}=\mathcal{Q}^{+}$ and (\ref%
{inv}) implies the vanishing of the corresponding eigenvalue of the transfer
matrix. This does not mean that the second linear independent solution $%
\mathcal{Q}^{-}$ does not exist, it simply states that the normalization
constant $\mathcal{N}_{\mu =1}$ in the definition (\ref{Qm}) of $Q^{-}\neq 
\mathcal{Q}^{-}$ is zero. Note also that (\ref{zeroPS}) applies to
non-degenerate states only, which decrease in number as $M\gg N$ due to the
loop algebra symmetry at roots of unity \cite{DFM}.\medskip

\begin{center}
\medskip 
\begin{tabular}{||c||c||c||c||c||}
\hline
$M$ & 3 & 5 & 7 & 9 \\ \hline
$N=3$ & 1/3 & 1/10 & 1/35 & 1/126 \\ \hline
$N=5$ & 3/3 & 8/10 & 21/35 & 55/126 \\ \hline
$N=7$ & 3/3 & 10/10 & 33/35 & 108/126 \\ \hline
\end{tabular}
\medskip
\end{center}

{\small \textbf{Table 1}. Shown are the number of \textquotedblleft
maximal\textquotedblright\ solutions to the Bethe equations (i.e. }$%
n_{+}=M/2-S^{z}$ {\small Bethe roots above and }$n_{-}=M-n_{+}${\small \
below the equator with }$S^{z}=1/2${\small ) over the dimension of the
spin-1/2 sector. The deformation parameter is chosen to be q~=~exp(2}$\pi i$%
{\small /\emph{N}). For \emph{N}=3,5 it has been checked that the number of
\textquotedblleft maximal Bethe states\textquotedblright\ matches the
dimension of the kernel of the auxiliary matrix.}\medskip

In fact, for $N=3$ in the spin-sector $S^{z}=\pm 1/2$ there is only one
state with the expected number of Bethe roots above and below the equator,
the groundstate. This is in agreement with the results in \cite{Odd}.
However, our starting point is different from the one in \cite{Odd}. Instead
of making a conjecture on the explicit form of the groundstate of the
six-vertex model, we simply start from the assumption that there exists an
eigenstate with $m=(M-1)/2$ Bethe roots in the spin $S^{z}=1/2$ sector.
According to (\ref{deg}) complete strings cannot be present and thus (\ref%
{PS}) must be a constant. But because of (\ref{zeroPS}) with $Q^{+}=\mathcal{%
Q}^{+}$ this constant is vanishing and we have the difference equation, 
\begin{equation}
(1-z)^{M}\mathcal{Q}%
^{+}(zq^{2})+q^{-1}(1-zq^{2})^{M}Q^{+}(zq^{-2})+q^{-2}(1-zq^{-2})^{M}Q^{+}(z)=0\ .
\label{diffN3}
\end{equation}
As our conventions differ from Stroganov's we review his calculation in the
appendix and show that (\ref{diffN3}) has a unique solution which can be
expressed in terms of hypergeometric functions. The same holds true for the
second linear independent solution $\mathcal{Q}^{-}$ which has $m+1$ roots.

In the case of general $N\in 2\mathbb{N}+1$ similar difference equations
follow. Let $M=2m+1,$ then the eigenvalues of singlet states with $n_{\infty
}=n_{S}=0$ in the spin $S^{z}=1/2$ sector with $N\geq 3$ satisfy 
\begin{equation*}
\sum_{\ell =1}^{N}q^{\mp \ell }f_{N}(zq^{2\ell })=0,\qquad
f_{N}(z)=(1-z)^{M}\prod_{\ell =1}^{N-2}\mathcal{Q}^{\pm }(zq^{2\ell })
\end{equation*}
implying the following sum rules in terms of the elementary symmetric
polynomials (\ref{epm}) in the $m$ Bethe roots above and the $m+1$ Bethe
roots below the equator, 
\begin{equation}
0=\sum_{k+l=n}\binom{M}{k}\sum_{k_{1}+\cdots
+k_{N-2}=l}~\tprod_{j=1}^{N-2}q^{-2j\,k_{j}}e_{k_{j}}^{\pm }\ .
\label{diffNe}
\end{equation}
Here the integer $n$ takes all values in the range 
\begin{equation}
0<n=\frac{N\pm 1}{2}\func{mod}N\leq N~\frac{M\mp 1}{2}\pm 1
\end{equation}
and the different summation variables run over the intervals, 
\begin{equation}
0\leq k\leq M,\quad \quad 0\leq k_{j}\leq \frac{M\mp 1}{2}\;.
\end{equation}
In general the set of equations (\ref{diffNe}) is of order $N-2$ and only
for $N=3$ becomes linear in the variables $\{e_{k}^{\pm }\}$, where the
equations are particularly simple to solve. Nevertheless, these sum rules
are still an advantage over the Bethe ansatz equations which are of order $M$%
.

\section{Conclusions}

Let us summarize the new results obtained for the six-vertex model at roots
of unity. First of all the discussion has been extended from even to odd
spin-chains exploiting that the construction procedure for the auxiliary
matrices in \cite{KQ} does not have the same limitations as the one in
Baxter's book \cite{BxBook}. This allowed us to reveal the major differences
in the spectrum of the six-vertex model at roots of unity between these two
cases.

\begin{enumerate}
\item When the length of the spin-chain is even there are degeneracies in
the spectrum of the transfer matrix for all roots of unity. These
degeneracies are reflected in the spectrum of the auxiliary matrices through
the occurrence of \textquotedblleft complete string
factors\textquotedblright , see (\ref{PS}) in the text. The number of these
strings, i.e. the degree $n_{S}$ of the polynomial $P_{S}$ in (\ref{QE}),
determines the degeneracy of the corresponding eigenspace of the transfer
matrix to be $2^{n_{S}}$. This is in accordance with the observations made
in \cite{DFM}. In order to arrive at this result we made use of the crucial
functional equation (\ref{QQQ}) which severely restricts the dependence of
the complete string factors on the free parameter $\mu $ entering the
definition of the auxiliary matrix (\ref{Qmu}). In addition, we employed (%
\ref{QQQ}) to prove the identity (\ref{PS}) which states that the string
factors are determined (up to their dependence on the aforementioned
parameter $\mu $) by the solution to the Bethe ansatz equations and the
number of infinite Bethe roots which fix the eigenvalue of the transfer
matrix. These results had previously been proved for $N=3$ only and
conjectured to hold true for $N>3$ \cite{KQ2}. Moreover, we deduced that the
Bethe roots appear twice in the eigenvalue of the auxiliary matrices, once
in the factor $P_{B}=Q^{+}$ and once multiplied by the factor $\mu ^{2}$ in
the factor $P_{\mu }$ of the eigenvalue (\ref{QE}). A second linear
independent solution to the $TQ$-equation was not found.

\item For spin-chains with an odd number of sites the novel feature was the
appearance of such a second linear independent solution to the $TQ$-equation
below the equator.

\qquad For primitive roots of unity of odd order this second solution does
not exist for all eigenstates of the transfer matrix, but only for those
which are singlets and have no infinite roots. For these eigenstates of the
transfer matrix we have shown that due to the functional equation (\ref{QQQ}%
) the corresponding eigenvalues of the auxiliary matrices must vanish, see
equation (\ref{zeroPS}) in the text. The number of these states, i.e. the
dimension of the kernel of the auxiliary matrix, will become smaller as the
length of the spin-chain starts to exceed the order of the root of unity,
i.e. $M\gg N$. (For $N=3$ we in particular saw that there is only one such
singlet state for all odd $M$ and it corresponds to Stroganov's solutions of
the $TQ$-equation for the groundstate of the $XXZ$ spin-chain at $\Delta
=-1/2$ \cite{Odd}.) This decrease in number can be understood in terms of
the loop algebra symmetry \cite{DFM} of the six-vertex transfer matrix. As
the length of the spin-chain $M$ grows more and more of the transfer matrix'
eigenstates organize into larger and larger multiplets spanning the
irreducible representations. Similar to the case of even spin-chains these
degenerate states within the multiplets give rise to complete strings in the
eigenvalues of the auxiliary matrices with the same formula $2^{n_{S}}$
yielding the multiplicity of the transfer matrix eigenvalue.

\qquad For primitive roots of unity of even order the solution below the
equator always exists, here, however, the eigenvalues of the transfer matrix
do not vanish. We showed the absence of infinite Bethe roots as well as
complete strings, leaving at most a double degeneracy in the spectrum of the
transfer matrix due to spin-reversal symmetry. The latter is broken by the
auxiliary matrices and we used this fact to identify $P_{B}=Q^{+}$ and $%
P_{\mu }=Q^{-}$ with the solutions to the $TQ$-equation above and below the
equator, respectively. The explicit construction of the $Q$-matrices in \cite%
{KQ}, see the definition (\ref{Qmu}) in this article, guarantees therefore
the existence of these two solutions, a fact implicitly assumed in \cite%
{PrSt} for the case of \textquotedblleft generic $q$\textquotedblright .
What is still lacking at the moment is an understanding of the physical
significance behind the existence of two linear independent solutions
opposed to the case when there is only one. We hope to address this question
in a future publication.
\end{enumerate}

In this article we have focussed on the case when $q$ is a root of unity to
discuss the spectra of the auxiliary matrices constructed in \cite{KQ}. But
as mentioned in the introduction analogous $Q$-operators have also been
constructed when $q$ is not a root of unity \cite{RW02}. Their spectra
together with the resolution of certain convergence problems originating
from an infinite-dimensional auxiliary space have been discussed in \cite%
{KQ3}. As explained therein one in general needs to impose quasi-periodic
boundary conditions on the lattice in order to obtain a well-defined
auxiliary matrix. For instance, in the critical regime $|q|~=1$ the twist
parameter $\lambda $ has to be of modulus smaller than one, $|\lambda |<1,$
to guarantee absolute convergence \cite{KQ3}. This twist parameter enters
the definition of the fusion matrices (\ref{Tn}) as 
\begin{equation}
T^{(n+1)}(zq^{-n-1})=\limfunc{Tr}_{0}\lambda ^{\pi ^{(n)}(h_{1})\otimes
1}L_{0M}^{(n+1)}(zq^{n+1})\cdots L_{01}^{(n+1)}(zq^{-n-1})
\end{equation}%
and modifies the Wronskian (\ref{wronski}) in the following manner, 
\begin{equation}
\lambda ^{-1}q^{S^{z}}\mathcal{Q}^{+}(zq^{2})\mathcal{Q}^{-}(z)-\lambda
q^{-S^{z}}\mathcal{Q}^{+}(z)\mathcal{Q}^{-}(zq^{2})=(\lambda
^{-1}q^{S^{z}}-\lambda q^{-S^{z}})(1-zq^{2})^{M}\ .
\end{equation}%
Numerical computations show that both solutions $\mathcal{Q}^{\pm }$ exist
for $M$ even and odd. Proceeding similar as we did for the case of odd roots
of unity we can iteratively solve this equation for, say $\mathcal{Q}^{+}$,
to obtain 
\begin{multline*}
\mathcal{Q}^{-}(z)=\lim_{n\rightarrow \infty }\lambda ^{2n}q^{-2nS^{z}}\frac{%
\mathcal{Q}^{+}(z)\mathcal{Q}^{-}(zq^{2n})}{\mathcal{Q}^{+}(zq^{2n})} \\
+(1-\lambda ^{2}q^{-2S^{z}})\mathcal{Q}^{+}(z)\sum_{\ell =0}^{\infty }\frac{%
\lambda ^{2\ell }q^{-2\ell S^{z}}(1-zq^{2\ell +2})^{M}}{\mathcal{Q}%
^{+}(zq^{2\ell +2})\mathcal{Q}^{+}(zq^{2\ell })}\ .
\end{multline*}%
In the limit $n\rightarrow \infty $ the first term on the right hand side
tends to zero as $|\lambda |<1$ and $|q|=1$. The above expression then
matches the results obtained for the eigenvalues of the $Q$-operator from
the algebraic Bethe ansatz, see equations (75-78) in \cite{KQ3}, 
\begin{equation*}
Q_{\leq }(z;r_{0},r_{1})=(-1)^{M}r_{0}^{-S^{z}}(1-\lambda
^{2}q^{-2S^{z}})Q^{+}(zr_{1})Q^{-}(z)\ .
\end{equation*}%
Here, up to some trivial normalization factors, we can identify $Q^{\pm
}\propto \mathcal{Q}^{\pm }$. Thus, similar as in the root-of-unity case the
spectrum of the auxiliary matrices \cite{RW02,KQ3} when $q$ is not a root of
unity\ contains both solutions to the $TQ$-equation, the one above and the
one below the equator. Moreover, these two solutions are related by the
analogue of the formula (\ref{inv}) where the summation extends now over an
infinite interval.

In the text we commented on a similarity between the relation (\ref{inv})
and a eight-vertex functional equation conjectured by Fabricius and McCoy 
\cite{FM8v,FM8v2} for Baxter's 1972 auxiliary matrix $Q_{\text{8v}}$ \cite%
{Bx72} at coupling values $\eta =mK/N^{\prime }$ (see equation (3.10) in 
\cite{FM8v} or (3.1) in \cite{FM8v2}), 
\begin{multline}
e^{-i\pi uM/2K}Q_{\text{8v}}(u-iK^{\prime })= \\
A~Q_{\text{8v}}(u)\sum_{\ell =0}^{N^{\prime }-1}\frac{h^{M}(u-(2\ell
+1)K/N^{\prime })}{Q_{\text{8v}}(u-2\ell K/N^{\prime })Q_{\text{8v}%
}(u-(2\ell +1)K/N^{\prime })}\ .
\end{multline}
Here $h(u)=\theta _{4}(0)\theta _{1}(\pi u/2K)\theta _{4}(\pi u/2K)$ in
terms of Jacobi's theta-functions with modular parameter $p=\exp (-\pi
K^{\prime }/K)$. We have made the replacements $L\rightarrow N^{\prime }$
(the order of the root of unity) and$\ N\rightarrow M$ (the length of the
spin-chain) in the notation of \cite{FM8v,FM8v2}. For general $N^{\prime }$
the explicit form of the constant $A$ is as yet unknown. This functional
equation has been proved for the free fermion case when $M$ is even and
numerically verified for coupling values corresponding to roots of unity of
order three, see the comment after (3.1) in \cite{FM8v2}.

If one identifies $Q_{\text{8v}}(u-iK^{\prime })$ with $Q^{-}(e^{u}q^{-1})$
and $Q_{\text{8v}}(u)$ with $Q^{+}(e^{u}q^{-1})$ in the six-vertex limit the
similarity becomes apparent. This is further supported by the observation
that the transformation $u\rightarrow u-iK^{\prime }$ corresponds to
spin-reversal in the eight-vertex Boltzmann weights\footnote{%
The author is thankful to Barry McCoy for discussions on this point.}. It is
tempting to speculate on further identities such as the elliptic analogue of
the relation (\ref{wronski3}) for instance. While these identifications can
be conjectured and numerically investigated on the level of eigenvalues, the
analogous construction of the eight-vertex $Q$-matrices corresponding to (%
\ref{Qmu}), which would allow one to prove existence, is a more complicated
problem.{\small \medskip }

\noindent \textbf{Acknowledgments}. It is a pleasure to thank Harry Braden
for comments and Barry McCoy for fruitful discussions. The author is also
grateful to the Research Institute for Mathematical Sciences, Kyoto
University, where part of this work has been carried out. This work has been
financially supported by the EPSRC Grant GR/R93773/01.

\appendix

\section{Derivation of the functional equation}

As mentioned earlier the proof of the functional equation (\ref{QQQ})
employs representation theory and is deduced from the following non-split
exact sequence describing the decomposition of the tensor product $\pi
_{w}^{\mu }\otimes \pi _{1}^{\nu }$, 
\begin{equation}
0\rightarrow \pi _{w^{\prime }}^{\mu ^{\prime }}\overset{\imath }{%
\hookrightarrow }\pi _{w}^{\mu }\otimes \pi _{1}^{\nu }\overset{\tau }{%
\rightarrow }\pi _{w^{\prime \prime }}^{\mu ^{\prime \prime }}\otimes \pi
_{z^{\prime }}^{(N^{\prime }-2)}\rightarrow 0\ .
\end{equation}
The various parameters appearing in the representations are not all
independent but satisfy the relations 
\begin{equation}
w=\mu \nu q^{2},\quad \mu ^{\prime }=\mu \nu q,\quad w^{\prime }=\mu q,\quad
\mu ^{\prime \prime }=\mu \nu q^{-N^{\prime }+1},\quad w^{\prime \prime
}=\mu q^{N^{\prime }+1},\quad z^{\prime }=\nu q^{N^{\prime }+1}\;.
\label{var}
\end{equation}
Notice that we have set the second evaluation parameter in the tensor
product $\pi _{w}^{\mu }\otimes \pi _{1}^{\nu }$ equal to one. The general
case $\pi _{w}^{\mu }\otimes \pi _{u}^{\nu }$ is obtained by simply
replacing $w\rightarrow wu,\,w^{\prime }\rightarrow w^{\prime }u$ and $%
w^{\prime \prime }\rightarrow w^{\prime \prime }u,\,z^{\prime }\rightarrow
z^{\prime }u$. The line of argument is analogous to the one applied in \cite%
{KQ} to derive the $TQ$-equation via (\ref{seq1}) and the proof of (\ref{QQQ}%
) in \cite{KQ2} for $N=3$, whence we will be rather brief in presenting the
various steps of the proof.

\paragraph{The inclusion.}

First we determine the subrepresentation $\pi _{w^{\prime }}^{\mu ^{\prime
}} $ contained in the tensor product $\pi _{w}^{\mu }\otimes \pi _{1}^{\nu }$
when $w$ is tuned to the special value given in (\ref{seq2}). This will
yield the first term on the right hand side of (\ref{QQQ}). The
corresponding inclusion map $\imath :\pi _{w^{\prime }}^{\mu ^{\prime
}}\hookrightarrow \pi _{w}^{\mu }\otimes \pi _{1}^{\nu }$ is determined by
identifying the lowest weight vectors in both representations, 
\begin{equation}
\pi _{w^{\prime }}^{\mu ^{\prime }}\ni \left\vert 0\right\rangle \overset{%
\imath }{\hookrightarrow }\left\vert 0\right\rangle \otimes \left\vert
0\right\rangle \in \pi _{w}^{\mu }\otimes \pi _{1}^{\nu }\;.
\end{equation}
The remaining relations for the rest of the vectors in the included
subrepresentation is obtained by successive action of the quantum group
generators via the formula 
\begin{equation*}
\pi _{w^{\prime }}^{\mu ^{\prime }}(x)\left\vert 0\right\rangle \overset{%
\imath }{\hookrightarrow }(\pi _{w}^{\mu }\otimes \pi _{1}^{\nu })\Delta
(x)\left\vert 0\right\rangle \otimes \left\vert 0\right\rangle \;.
\end{equation*}
The above formula suffices to compute the various parameters. For instance,
the parameter $\mu ^{\prime }$ in (\ref{var}) is computed from the action of
the Cartan element $x=q^{h_{1}}$. The evaluation parameters $w,w^{\prime }$
are deduced as follows. First act with $f_{1}$ on the lowest weight vector
to obtain 
\begin{equation*}
\pi _{w^{\prime }}^{\mu ^{\prime }}(f_{1})\left\vert 0\right\rangle =\overset%
{\imath }{\,\left\vert 1\right\rangle \hookrightarrow }(\pi _{w}^{\mu
}\otimes \pi _{1}^{\nu })\Delta (f_{1})\left\vert 0\right\rangle \otimes
\left\vert 0\right\rangle =\nu q\left\vert 1\right\rangle \otimes \left\vert
0\right\rangle +\left\vert 0\right\rangle \otimes \left\vert 1\right\rangle
\;.
\end{equation*}
Alternatively, one obtains via the generator $e_{0}$, 
\begin{equation*}
\pi _{w^{\prime }}^{\mu ^{\prime }}(e_{0})\left\vert 0\right\rangle
=w^{\prime }\overset{\imath }{\left\vert 1\right\rangle \hookrightarrow }%
(\pi _{w}^{\mu }\otimes \pi _{1}^{\nu })\Delta (e_{0})\left\vert
0\right\rangle \otimes \left\vert 0\right\rangle =w\left\vert 1\right\rangle
\otimes \left\vert 0\right\rangle +\mu q\left\vert 0\right\rangle \otimes
\left\vert 1\right\rangle \;.
\end{equation*}
Comparing both results leads to the stated values of $w,w^{\prime }$ in (\ref%
{var}). The scalar coefficient in front of the first term on the right hand
side of the functional equation (\ref{QQQ}) is obtained from the identity 
\begin{equation}
L_{13}^{\mu }(z\mu q^{2})L_{23}^{\nu }(z/\nu )(\imath \otimes
1)=(zq^{2}-1)(\imath \otimes 1)L^{\mu \nu q}(z\mu q/\nu )
\end{equation}
which can be easily verified by acting on the lowest weight vector.

\paragraph{The projection.}

In order to compute the second term of the functional equation (\ref{QQQ})
we need to determine the representation in the quotient space $\pi _{w}^{\mu
}\otimes \pi _{1}^{\nu }/\pi _{w^{\prime }}^{\mu ^{\prime }}$ which is
projected out by the map $\tau $. One proceeds in an analogous manner. The
projection map $\tau $ is fixed by identifying this time the highest weight
vector in both representation spaces, 
\begin{equation}
\pi _{w}^{\mu }\otimes \pi _{1}^{\nu }\ni \left\vert N^{\prime
}-1\right\rangle \otimes \left\vert N^{\prime }-1\right\rangle \overset{\tau 
}{\rightarrow }\left\vert N^{\prime }-1\right\rangle \otimes \left\vert
N^{\prime }-2\right\rangle \in \pi _{w^{\prime \prime }}^{\mu ^{\prime
\prime }}\otimes \pi _{z^{\prime }}^{(N^{\prime }-2)}\;.
\end{equation}
Again the value of $\mu ^{\prime \prime }$ can be inferred from the action
of the Cartan element $x=q^{h_{1}}$ via the formula 
\begin{equation*}
(\pi _{w}^{\mu }\otimes \pi _{1}^{\nu })\Delta (x)\left\vert N^{\prime
}-1\right\rangle \otimes \left\vert N^{\prime }-1\right\rangle \overset{\tau 
}{\rightarrow }(\pi _{w^{\prime \prime }}^{\mu ^{\prime \prime }}\otimes \pi
_{z^{\prime }}^{(N^{\prime }-2)})\Delta (x)\left\vert N^{\prime
}-1\right\rangle \otimes \left\vert N^{\prime }-2\right\rangle \;.
\end{equation*}
Setting $x=f_{1}f_{0}$ and $x=e_{0}e_{1}$ one obtains the desired evaluation
parameters $w^{\prime \prime }$ and $z^{\prime }$ detailed in (\ref{var}).
For instance, from the left hand side of the above equation one obtains 
\begin{multline*}
(q-q^{-1})^{2}(\pi _{w}^{\mu }\otimes \pi _{1}^{\nu })\Delta
(f_{1}f_{0})\left\vert N^{\prime }-1\right\rangle \otimes \left\vert
N^{\prime }-1\right\rangle = \\
\left\{ \frac{\mu +\mu ^{-1}-\mu q^{-2}-\mu ^{-1}q^{2}}{w}+\nu +\nu
^{-1}-\nu q^{-2}-\nu ^{-1}q^{2}\right\} \left\vert N^{\prime
}-1\right\rangle \otimes \left\vert N^{\prime }-1\right\rangle
\end{multline*}
while the right hand side is computed to 
\begin{multline*}
(q-q^{-1})^{2}(\pi _{w^{\prime \prime }}^{\mu ^{\prime \prime }}\otimes \pi
_{z^{\prime }}^{(N^{\prime }-2)})\Delta (f_{1}f_{0})\left\vert N^{\prime
}-1\right\rangle \otimes \left\vert N^{\prime }-2\right\rangle = \\
\frac{\mu \nu q^{-N^{\prime }+1}+(\mu \nu )^{-1}q^{N^{\prime }-1}-\mu \nu
q^{-N^{\prime }-1}-(\mu \nu )^{-1}q^{N^{\prime }+1}}{w^{\prime \prime }}%
\left\vert N^{\prime }-1\right\rangle \otimes \left\vert N^{\prime
}-2\right\rangle \\
+\frac{(q-q^{-1})^{2}[N^{\prime }-2]_{q}}{z^{\prime }}\left\vert N^{\prime
}-1\right\rangle \otimes \left\vert N^{\prime }-2\right\rangle \ .
\end{multline*}
Matching the coefficients in both results yields the stated expressions for
the parameters. In the case of the quotient projection there is only a
trivial additional scalar factor as we have the equality 
\begin{equation}
(\tau \otimes 1)L_{13}^{\mu }(z\mu q^{2})L_{23}^{\nu }(z/\nu )=q^{N^{\prime
}}L_{13}^{\mu \nu q^{-N^{\prime }+1}}(z\mu q^{-N^{\prime }+1}/\nu
)L_{23}^{(N^{\prime }-2)}(zq^{N^{\prime }+1})(\tau \otimes 1)\ .
\end{equation}
Again this is most easily calculated by acting with both sides of the
equation on the highest weight vector. This completes the proof of the
functional equation.

\section{Stroganov's solution revisited}

As explained in the text the assumption that there exists an eigenstate with 
$m=(M-1)/2$ Bethe roots in the spin $S^{z}=1/2$ sector implies via (\ref{deg}%
), (\ref{PS}) and (\ref{zeroPS}) with $Q^{+}=\mathcal{Q}^{+}$ that we have
the difference equation, 
\begin{equation}
(1-z)^{M}Q^{+}(zq^{2})+q^{-1}(1-zq^{2})^{M}Q^{+}(zq^{-2})+q^{-2}(1-zq^{-2})^{M}Q^{+}(z)=0\ .
\end{equation}
Expanding 
\begin{equation*}
(1-z)^{M}Q^{+}(zq^{2})=1+\sum_{n=1}^{3m+1}c_{n}z^{n}
\end{equation*}
we infer that the difference equation implies 
\begin{equation*}
c_{n}=0\quad \text{if\quad }n=2\func{mod}3\;.
\end{equation*}
The remaining coefficients can be determined from the fact that $%
(1-z)^{M}Q^{+}(zq^{2})$ has an $M$-fold zero at $z=1$. Applying the method
of Lagrange interpolating polynomials, similar as it has been done in \cite%
{ODEXXZ,Odd}, one finds the ratios 
\begin{equation*}
\frac{c_{3n+3}}{c_{3n}}=\frac{(n-m)(n-m-1/3)}{(n+1)(n+2/3)}\quad \quad \text{%
and\quad \quad }\frac{c_{3n+4}}{c_{3n+1}}=\frac{(n-m)(n+1/3-m)}{(n+1)(n+4/3)}
\end{equation*}
together with 
\begin{equation*}
c_{0}=1\quad \quad \text{and\quad \quad }c_{1}=-(4/3)_{m}/(2/3)_{m}\ .
\end{equation*}
Here $(x)_{m}$ is the Pochhammer symbol. From the ratios of the coefficients
we infer that there is a unique solution which can be expressed in terms of
hypergeometric functions 
\begin{equation}
(1-z)^{M}Q^{+}(zq^{2})=\,_{2}F_{1}(-m,-m-\tfrac{1}{3},\tfrac{2}{3};z^{3})-%
\frac{(\frac{4}{3})_{m}}{(\frac{2}{3})_{m}}\;z\,_{2}F_{1}(\tfrac{1}{3}-m,-m,%
\tfrac{4}{3};z^{3})\ .
\end{equation}
Note that this solution is not simply obtained by multiplying Stroganov's
solution (11) in \cite{Odd} with an exponential factor. This is due to the
fact that we solved the difference equation (\ref{diffN3}) in terms of
polynomials which are regular at origin, while Stroganov's solution applies
to Laurent series.

\paragraph{Second solution.}

Besides the solution for the Bethe polynomial we just obtained, there is a
second solution \textquotedblleft beyond the equator\textquotedblright\ as
it possesses $m+1$ roots. Set 
\begin{equation*}
(1-z)^{M}\mathcal{Q}^{-}(zq^{2})=1+\sum_{n=1}^{3m+2}c_{n}^{\prime }z^{n}
\end{equation*}
then it obeys the difference equation with $S^{z}=-1/2$, i.e. 
\begin{equation*}
(1-z)^{M}\mathcal{Q}^{-}(zq^{2})+q(1-zq^{2})^{M}\mathcal{Q}%
^{-}(zq^{-2})+q^{2}(1-zq^{-2})^{M}\mathcal{Q}^{-}(z)=0\ .
\end{equation*}
This implies for the coefficients 
\begin{equation*}
c_{n}^{\prime }=0\quad \text{if\quad }n=1\func{mod}3\;.
\end{equation*}
As before the solution to this set of equations can be expressed in terms of
hypergeometric functions, 
\begin{equation}
(1-z)^{M}\mathcal{Q}^{-}(zq^{2})=\,_{2}F_{1}(-m,-m-\tfrac{2}{3},\tfrac{1}{3}%
;z^{3})-\frac{(\frac{5}{3})_{m}}{(\frac{1}{3})_{m}}\;z^{2}\,_{2}F_{1}(\tfrac{%
2}{3}-m,-m,\tfrac{5}{3};z^{3})\ .
\end{equation}

\paragraph{Groundstate eigenvalue.}

From the difference equation it follows that the eigenvalue of the transfer
matrix is given by 
\begin{equation*}
T(z)P_{B}=q^{\pm \frac{1}{2}}(z-1)^{M}\frac{\mathcal{Q}^{\pm }(zq^{2})}{%
\mathcal{Q}^{\pm }(z)}+q^{\mp \frac{1}{2}}(zq^{2}-1)^{M}\frac{\mathcal{Q}%
^{\pm }(zq^{-12})}{\mathcal{Q}^{\pm }(z)}=(zq^{-2}-1)^{M}
\end{equation*}
which matches the conjecture \cite{Bx72a,Bx89,ABB88} employed in \cite{Odd}.
The corresponding groundstate eigenvalue of the Hamiltonian is given by $H_{%
\text{XXZ}}=-M\;.$

\end{document}